# Antiferro- and Meta-magnetism in the *S*=7/2 Hollandite Analog EuGa$_2$Sb$_2$

Tanya Berry[1,2*], Sean R. Parkin,[3] Tyrel M. McQueen[1,2,4*]

1. Department of Chemistry, The Johns Hopkins University, Baltimore, Maryland 21218, USA
2. Institute for Quantum Matter, Department of Physics and Astronomy, The Johns Hopkins University, Baltimore, Maryland 21218, USA
3. Department of Chemistry, University of Kentucky, Lexington, Kentucky 40506, USA
4. Department of Materials Science and Engineering, The Johns Hopkins University, Baltimore, Maryland 21218, USA
* tberry9@jhu.edu and mcqueen@jhu.edu

**Abstract:**
Recent work analyzing the impact of non-symmorphic symmetries on electronic states has given rise to the discovery of multiple types of topological matter. Here we report the single crystal synthesis and magnetic properties of EuGa$_2$Sb$_2$, an Eu-based antiferromagnet structurally consisting of pseudo-1D chains of Eu ions related by a non-symmorphic glide plane. We find the onset of antiferromagnetic order at $T_N$ = 8 K. Above T$_N$ the magnetic susceptibility is isotropic. Curie-Weiss analysis suggests competing ferromagnetic and antiferromagnetic interactions, with $p_{eff}$ = 8.1$\mu_B$ as expected for 4f$^7$ $J = S = 7/2$ Eu$^{2+}$ ions. Below $T_N$ and at low applied magnetic fields, an anisotropy develops linearly, reaching $\chi_\perp/\chi_\parallel = 6$ at $T$ = 2 K. There is concomitant metamagnetic behavior along $\chi_\parallel$, with a magnetic field of $\mu_0 H \approx 0.5$ T sufficient to suppress the anisotropy. Independent of crystal orientation, there is a continuous evolution to a field polarized paramagnetic state with $M = 7\mu_B/$Eu$^{2+}$ at $\mu_0 H = 2$ T as $T \rightarrow 0$ K. Specific heat measurements show a recovered magnetic entropy of $\Delta S_{mag} \approx 16.4$ J.mol$^{-1}$.K$^{-1}$ from $T \sim 0$ K to $T = T_N$, close to the expected value of $R\ln(8)$ for an $S = 7/2$ ion, indicating negligible low dimensional spin fluctuations above $T_N$. We find no evidence of unusual behaviors arising either from the dimensionality or the presence of the non-symmorphic symmetries.

## I. Introduction:

Zintl phase compounds have earned a lot of recognition due to their polyanionic and cationic networks in areas of thermoelectric, topological materials, and magnetic materials.[1-10] The structural motifs offered by Zintl phase materials are quite appealing due to the electrostatic interactions between cations and anions and covalent interactions in the polyanionic framework. This setting causes the materials to often behave as two independent subunits, with separate magnetic and electronic behaviors from the cations and polyanionic framework if the cations are magnetic and the anions are non-magnetic.

When the cation is magnetic, the interplay between magnetism and structure could yield versatile ground states. Rare earth based Zintl materials is extensively studied in terms of electronic transport. However, their magnetic structures have scarcely been studied.[11-15]

Similarly, recent work has demonstrated how non-symmorphic symmetries can conspire to generate unique topological electronic states of matter, such as Dirac 6-fold and 8-fold fermions, as well as more general topological states such as those found in GdSb$_x$Te$_{2-x-\delta}$, CeSbTe, and GdSbTe.[16-22]

The magnetic Zintl phase EuGa$_2$Sb$_2$ was recently reported to exist based on a polycrystalline synthesis.[23] The reported crystal structure has space group *Pnma* and has non-symmorphic symmetries that relate to adjacent magnetic ions. Further, electron counting implies divalent Eu$^{2+}$, which has a half-filled f-shell (4f$^7$) and L=0, which considerably simplifies the interpretation of magnetic behavior. Thus, EuGa$_2$Sb$_2$ provides an opportunity to explore the effects of non-symmorphic symmetries on magnetic order.

This work reports the synthesis of single crystals of EuGa$_2$Sb$_2$ via flux growth, confirms the crystal structure by single crystal X-ray diffraction and studies the magnetic properties via magnetization and specific heat measurements. The crystal structure is describable as an analog of Hollandite, with a Ga$_2$Sb$_2$ framework forming 1D channels in which Eu ions form pseudo-1D chains. We find the onset of antiferromagnetic order at T$_N$ = 8 K. Above T$_N$, Curie-Weiss analysis yields $p_{eff}$ = 8.1$\mu_B$ as expected for 4f$^7$ $J = S = 7/2$ Eu$^{2+}$ ions. A positive Weiss temperature $\theta = 5.89$ K suggests competing ferromagnetic and antiferromagnetic interactions. The magnetic susceptibility is direction-independent above T$_N$, with an anisotropy developing linearly below T$_N$, reaching $\chi_\perp/\chi_\parallel = 6$ at T = 2 K. There is concomitant metamagnetic behavior along $\chi_\parallel$, with a magnetic field of $\mu_0 H \approx 0.5$ T sufficient to suppress the anisotropy. There is also a continuous evolution to a field polarized paramagnetic state with $M = 7\mu_B$/Eu$^{2+}$ at $\mu_0 H = 2$ T as $T \rightarrow 0$ K. Specific heat measurements show a sharp transition at T$_N$, with a broad tail of entropy loss at lower temperatures. The estimated recovered magnetic entropy of $\Delta S_{mag}$ = 16.4 J.mol$^{-1}$.K$^{-1}$ from $T \sim 0$ K to $T = T_N$ is close to the expected value of $R$ln(8) for an $S = 7/2$ ion. This behavior indicates little low dimensional spin fluctuations above T$_N$, despite the low dimensionality implied by the crystal structure, and attributed to the large value of *S*. Together; these results allow us to gain insights into the nature of magnetism in Eu-based Zintl compounds and its correlation to the structure and thermodynamics quantities.

## II. Methods:
### A. Experimental:

Single crystals of EuGa$_2$Sb$_2$ were synthesized from Eu (ingot, Yeemeida Technology Co., LTD 99.995%), Ga (ingot, Noah Tech 99.99%), and Sb (BTC, 99.999%) using the binary flux technique. The elements were put in Canfield crucibles (size: 2mL) in 50:110:150 ratio for Eu:Ga:Sb with a total composition mass of 5 grams. Ga and Sb were placed in the crucible at atmospheric conditions, while Eu was added last in an Ar-filled glove box. The Canfield crucible was placed in a quartz ampoule with quartz wool below and above the crucible, evacuated, and sealed under $1.2 \times 10^{-2}$ torr of pressure.[24] The evacuated ampoules were loaded in a box furnace at an angle of 45º. The temperature was ramped at 80 ºC/h to *T*=550 ºC for 12 hours. This step allows for Ga and Sb binary flux to be in a liquid state. The furnace was then ramped from *T*=550°C to *T*=1100°C at the rate of 80 ºC/h and held for 24 hr. The furnace was then slowly cooled to *T*=650 ºC at the rate of 5 ºC/h, then removed hot, inverted, and immediately centrifuged. Centrifugation took 2-3 minutes. Rod shaped crystals of size 1-1.5 mm along the long direction were removed from the frit. These single crystals of EuGa$_2$Sb$_2$ were found to be stable on the benchtop.

Powder X-ray diffraction (XRD) patterns were collected over an angle range of 5-60º on a laboratory Bruker D8 Venture Focus diffractometer that utilizes a LynxEye detector and Cu Kα radiation. Structural refinements were performed with GSAS-II, and resulting structures were visualized with Vesta software.[25]

Single crystal X-ray data were collected on a Bruker-Nonius X8 Proteum (Mo Kα radiation) diffractometer. SADABS was used to apply the absorption correction using the 'multi-scan' approach.[25] All calculations were performed using the SHELX software package. The structures were solved by direct methods, and successive interpretations of difference Fourier maps were followed by least-squares refinement.[26,27]

Magnetization data were collected on a Quantum Design Magnetic Property Measurement System (MPMS). Magnetic susceptibility was approximated as magnetization divided by the applied magnetic field ($\chi \approx M/H$). In addition, heat capacity data were collected on a Quantum Design Physical Properties Measurement System using the semi-adiabatic method and a 1% temperature rise.

## III. Result and Discussion:

### A. Structure

Powder XRD scans of $EuGa_2Sb_2$ were consistent with the previously reported structure in space group Pnma (62). The as grown single crystals of $EuGa_2Sb_2$ showed phase purity via powder refinement. The results of single crystal diffraction data refinements, tables 1 and 2, are consistent with the prior literature reports from powder diffraction refinements. From an electron counting perspective, the compound is expected to be built of $Eu^{2+}$ cations, and a $[Ga_2Sb_2]^{2-}$ anionic Zintl framework, and that is indeed the case. Figure 1(a) shows that the structure is built of $[Ga_2Sb_2]^{2-}$ units that form structural motifs of tetrahedral and bridging Sb in a square-chain ladder format. The squares consist of 2 bridging Sb and 2 Ga atoms, and the ladder consisting of Ga-Ga chains. These form an extended network with one-dimensional channels in which $Eu^{2+}$ ions reside.

The $EuGa_2Sb_2$ structure is comparable to $EuLu_2Se_4$ and $K_2Sn_3O_7$, crystallizing in the Pnma space group and analogous to the Hollandite structure type.[28,29] In Figure-1(c) and (d), we see the classic α-$MnO_2$ structure type of Hollandite in $EuLu_2Se_4$ and $K_2Sn_3O_7$. In these structures, there is a presence of 1D chains with the cations $Eu^{2+}$ and $K^+$ ions spaced 4.0470(4) Å and 3.12250(9) Å apart, respectively, within each chain. In comparison, the structure of $EuGa_2Sb_2$ is an analog with $Eu^{2+}$ as the cation forming chains with a spacing of 4.3225(8) Å. One difference is that the anion chains of $[Ga_2Sb_2]^{2-}$ form smaller networks of tetrahedral and bridging Sb in a square-chain ladder format instead of face sharing octahedral in the other two structures. This pseudo-Hollandite structure in $EuGa_2Sb_2$ is also structurally distinct from other pseudo-Hollandite structures such as $PbIr_4Se_8$ and $TlCr_5Se_8$ occupancy of the 1D chains with transition metals and charge balancing of $Tl^{1+}$ and $Cr^{3+}$.[30,31] In all cases there are non-symmorphic glide plane symmetries present that relate $Eu^{2+}$ ions in adjacent channels.

### B. Heat capacity

Temperature-dependent heat capacity measurements were carried out to understand the magnetic contribution of Eu and phononic contribution in the $EuGa_2Sb_2$ single crystals. Figure 2 shows the heat capacity as a function of temperature in $EuGa_2Sb_2$ from 2-225 K. The heat capacity plot showed a sharp phase transition at $T = 7.4$ K. At $T = 225$ K, the Dulong-Petit limit was observed. This limit is calculated using the $C_p = 3NR$, where $N$ is the number of atoms and R is the ideal gas constant.[32] In $EuGa_2Sb_2$, the number of atoms is 5, and the value is approximately equal to 124 J $mol^{-1}$ $K^{-1}$ respectively. In order to generate the change in entropy from the magnetic order, the phonons were subtracted by modeling the high temperature specific heat using a two Debye phonon model and then removing that as the

phonon contribution at all temperatures. The two Debye model is:

$$\frac{C_p}{T} = \frac{C_D(\theta_{D1}, s_1, T)}{T} + \frac{C_D(\theta_{D2}, s_2, T)}{T} \quad (1)$$

$$C_D(\theta_D, T) = 9sR\left(\frac{T}{\theta_D}\right)^3 \int_0^{\theta_D/T} \frac{(\theta/T)^4 e^{\theta/T}}{[e^{\theta/T}-1]^2} d\frac{\theta}{T} \quad (2)$$

Where $\theta_{D1}$ and $\theta_{D2}$ are the Debye temperatures, $s_1$ and $s_2$ are the oscillator strengths, and R is the molar Boltzmann constant. The model parameters from the least-squares refinement to the data for $T > 16$ K, Figure 3(a), are given in Table 3. The total oscillator strength $s_1+s_2 = 5.2(2)$. This is in good agreement with the expected value of $1+2+2 = 5$, the total number of atoms per formula unit in $EuGa_2Sb_2$.

After subtracting this phonon contribution, the sample heat capacity from $T$=2-225K was integrated to determine the change in entropy corresponding to magnetic order in $EuGa_2Sb_2$. Figure 3(b) shows that the change in the magnetic entropy reaches a maximum of $\Delta S_{mag} \approx 16.4$ J.mol$^{-1}$.K$^{-1}$, close to the Rln(8) = 17.2 J.mol$^{-1}$.K$^{-1}$ expected for an $S$=7/2 system. The small discrepancy is attributable to the use of a linear extrapolation to capture the entropy from T = 0 to 2 K. The recovery of a full $R$ln(8) entropy is not unexpected, being seen in many divalent (Eu$^{2+}$) materials such as $EuMg_2Bi_2$ and $Eu_3In_2P_4$. This behavior contrasts to mixed valent (Eu$^{2+}$ and Eu$^{3+}$) compounds such as $EuZnSb_2$ and $EuIr_2In_8$ and supports our assignment of Eu$^{2+}$ on crystal-chemical grounds.[33-36] Further, recovery of the full Rln(8) indicates minimal splitting of states due to crystal field effects. This effect arises from the fact that L=0 and that the local point symmetry of each Eu$^{2+}$ ion is C$_m$: by symmetry, the ground state $^8S_{j=7/2}$ is allowed to split into four doublets, and the entropy from all four of these doublets must be recovered to reach Rln(8). However, the existence of these doublets may explain why the transition observed in specific heat is not a single sharp anomaly but instead has a pronounced tail extended well below the transition temperature. Third, more than 90% of the entropy is recovered below $T_N$. This trend indicates negligible loss of entropy above $T_N$, and implies a lack of low dimensional but longer range magnetic correlations. Despite the structure being built of pseudo-1D chains, the physical behavior does not have a regime in which the expected 1D physics is dominant; this might be due to the large $S$, which is known to suppress fluctuations.

## C. Magnetization

To further elucidate the magnetic properties of EuGa$_2$Sb$_2$, magnetization as a function of temperature and field were studied. The $M(T)$ plots in Figure 4(a) and (b) show an evident antiferromagnetic phase transition at $T_N$=8.3K and $T_N$=7.9K in the parallel and perpendicular direction to the b axis, respectively. To further quantify the magnetic susceptibility results, the high temperature, $T$=50-300K data were fitted to the Curie-Weiss law:

$$\chi = \frac{C}{T - \theta_{cw}} \quad (3)$$

Here C is the Curie constant, and $\theta_{CW}$ is the Weiss temperature. The fit is shown in Figure-4(d) and 4(e), respectively. The $p_{eff}$ extracted from the Curie constants are 8.16 and 8.15, respectively, close to the theoretical $p_{eff}$ for Eu$^{2+}$, in agreement with expectations and the specific heat results. Thus, the positive Weiss constants indicate dominant ferromagnetic interactions; in combination with the observed antiferromagnetic order, this indicates substantial antiferromagnetic and ferromagnetic interactions.

The $M/H$ behavior is found to vary with the strength of the applied magnetic fields. The field dependence on $M(T)$ shows a decrease in the ordering temperature with increasing field at low temperatures, becoming field independent beyond $T$=40K in both directions. Another observation was that in $\mu_oH\perp$b, the $\mu_oH$=0.5-2T range had the same magnitude in magnetization at $T$=2K. However, the $\mu_oH$//b showed a monotonic decrease. These observations indicate the presence of anisotropy at low temperatures. Figure 4(f) shows the magnetization ratio in the parallel and perpendicular directions; above $T_N$, the ratio is 1, indicating no anisotropy. Below $T_N$, the anisotropy rises linearly, reaching a value of 6 at $T$ = 2 K.

To further investigate the effect of anisotropy between the $\mu_oH$//b and $\mu_oH\perp$b in EuGa$_2$Sb$_2$ single crystals, magnetization as a function of the magnetic field was studied. Figure 5 (a) and (b) show that the effective moment reaches the theoretical saturation for Eu$^{2+}$ in parallel and perpendicular directions along the b plane at high fields, indicative of a field-polarized state. This result further confirms the assignment of Eu valence as Eu$^{2+}$. Further, the shapes of the $M(H)$ curves are different between the two directions below $T_N$, consistent with the observed appearance of anisotropy.

The measurements with $\mu_oH$//b show a metamagnetic transition seen at temperatures below the ordering temperature at $\mu_oH$=0.5T, as seen in Figure 5 (a) inset. Metamagnetism is seen in most Co and Eu compounds because of single-ion anisotropy, but not expected here due to the isotropic nature of a 4f$^7$ ($S$=7/2) state.[37,38] Instead, we attribute this metamagnetism as a consequence of the magnetic order, which fixes the spins relative to each other, making some directions more easily polarizable than others. These behaviors can be further observed in the derivatives: Figure 5 (c) and (d) show the change in derivative as a function of the field from the $M(H)$ plots, which further illustrate the metamagnetic transitions in $\mu_oH$//b below $T_N$; while $\mu_oH\perp$b has a monotonic trend. In general, these magnetism trends explain that the $\mu_oH$//b is the hard axis.

Combining these results yields the magnetic phase diagrams shown in Figure 6(a) and (b). Below ~0.5 T, the system is in an antiferromagnetic state in which the applied field is not sufficient to overcome the anisotropy. Between ~0.5 and ~2 T, the spins reorient following the applied field and develop increasing (magnetic) polarization until, finally, a field-polarized ferromagnetic/paramagnetic state is reached.

From these results, we can speculate on the type of magnetic order that is present. First, the material is antiferromagnetic; this necessitates antiferromagnetic order along at least one crystal direction. Second, the positive Weiss constant suggests ferromagnetic order along at least one direction. This result leaves three possible general states, shown in Figure-6 (c,d,e): ferromagnetic chains with alternating (c) or block (d) antiferromagnetic order between chains, or (e) antiferromagnetic chains with ferromagnetic order between chains. Finally, the direction of the magnetic moments is only loosely constrained: from the previous discussion in the $M(H)$ plots, it is clear that $\mu_oH$//b is the hard axis, but that can be explained either by spins that lie parallel to a with antiferromagnetic order perpendicular or by spins perpendicular to a but with ferromagnetic order parallel. Future neutron studies, challenging due to the highly absorbing nature of Eu, would be needed to discriminate between these possibilities.

**IV. Conclusion:**

The structure of EuGa$_2$Sb$_2$ crystallizes in the space group Pnma (62). The single crystals are synthesized for the first time using the flux technique. The EuGa$_2$Sb$_2$ structure is characterized to be a Hollandite analog and thus having pseudo-1D chains. The change in magnetic entropy recovered reaches the theoretical limit of $R$ln(8) for Eu$^{2+}$ ions, implying no frustration or fluctuations above T$_N$, and a small splitting in the crystal field levels. The magnetic properties reveal that EuGa$_2$Sb$_2$ is anisotropic below T$_N$, and has a metamagnetic transition along $\mu_oH$//b. The magnetic phase diagram reveals the two magnetic phases in the structure and three possibilities in the magnetic structure. Overall, our results help us gain insights into the structure and magnetism in EuGa$_2$Sb$_2$. For future work, neutron diffraction could further help elucidate the magnetic structure, and determine whether nonsymmorphic symmetries have any impact on the magnetic excitations that are present.

**Acknowledgments:**
This work was supported as part of the Institute for Quantum Matter and Energy Frontier Research Center, funded by the U.S. Department of Energy, Office of Science, Office of Basic Energy Sciences, under Award DE-SC0019331. The MPMS was funded by the National Science Foundation, Division of Materials Research, Major Research Instrumentation Program, under Award 1828490.

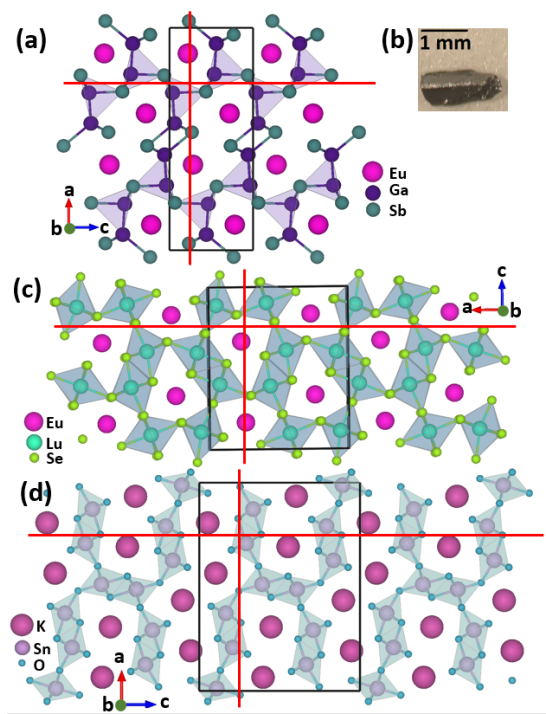

**Figure-1** Structure of (a) EuGa$_2$Sb$_2$, (c) EuLu$_2$Se$_4$, and (d) K$_2$Sn$_3$O$_7$ that crystallize in Pnma and are Hollandite-like structures containing 1D chains of ions in channels; red lines indicate the non-symmorphic glide planes. (b) Single crystal image of as grown EuGa$_2$Sb$_2$.

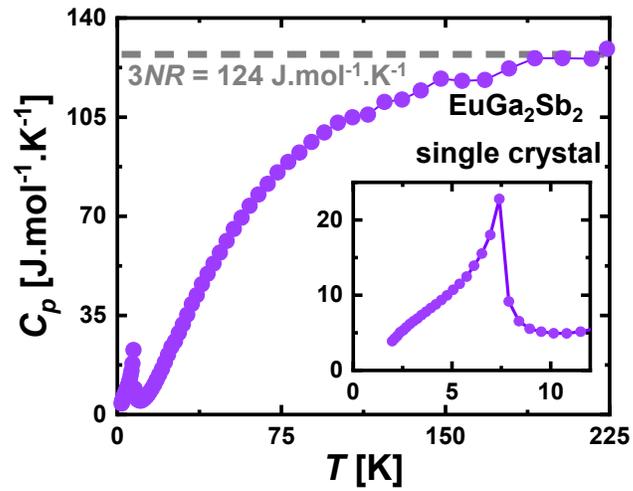

**Figure-2** Temperature-dependent heat capacity of EuGa$_2$Sb$_2$ single crystals at $\mu_0H$=0.1T from $T$=2-225K. The sharp transition at T=7K is attributed to the antiferromagnetic phase transition. The Dulong-Petit theoretical value of 3NR=124 J.mol$^{-1}$.K$^{-1}$ is reached around T = 200 K.

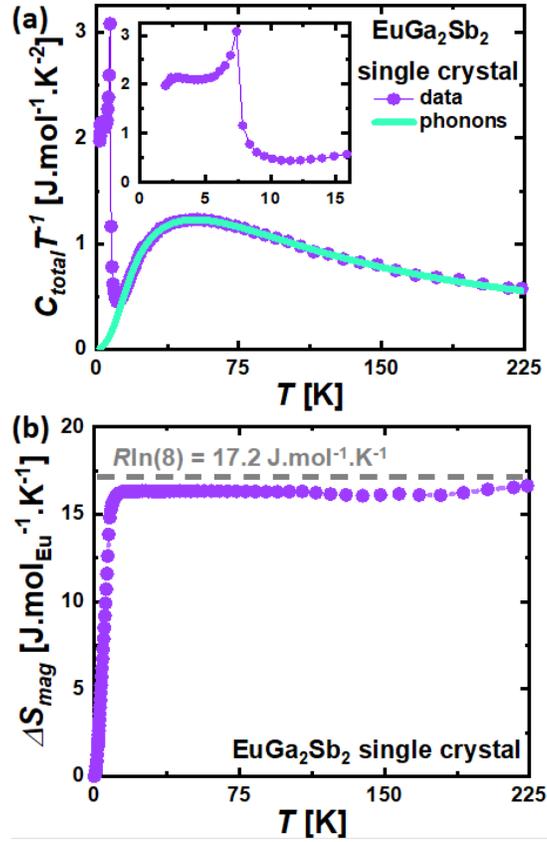

**Figure-3 (a)** Heat capacity divided by temperature as a temperature-dependent function of EuGa$_2$Sb$_2$ single crystals at $\mu_0H$=0.1T from $T$=2-225K. The purple is the data points, and the aqua is the phonons modeled using the two Debye models from $T$=16-225K. The sharp transition at $T$=7K is attributed to the antiferromagnetic phase transition. **(b)** The change in magnetic entropy was integrated after subtracting the phonons from 2-225K. The $\Delta S_{mag}$ is close to the $R\ln(8)$ expected for an L=S=7/2 system.

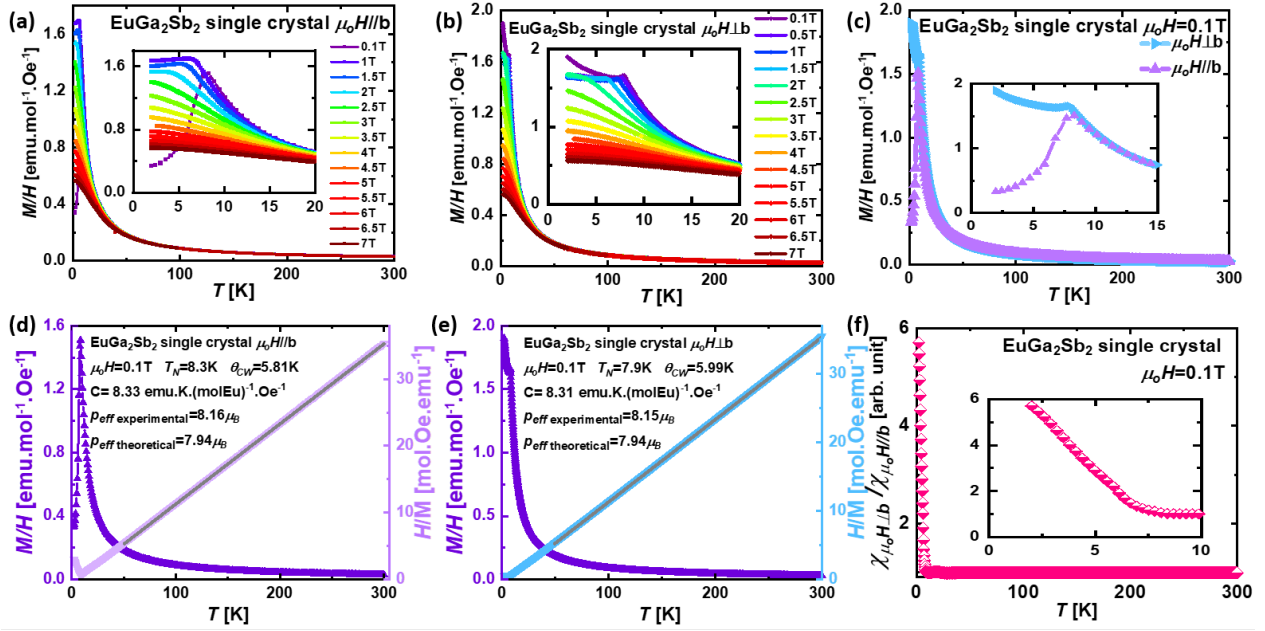

**Figure-4 (a)** Magnetization as a function of temperature with $\mu_oH//b$ and $\mu_oH$ = 0.1-7T and T=2-300K. The $\mu_oH$ = 0.1T data shows a clear AFM transition at T = 7 K, with a decrease in both sharpness of the transition and the temperature of the transition as the field increases. **(b)** Magnetization as a function of temperature $\mu_oH \perp b$ from $\mu_oH$ = 0.1-7T and T=2-300K. The $\mu_oH$ = 0.1T data shows a kink at T = 7K followed by an upturn, both of which are suppressed for $\mu_oH$ > 0.1T. **(c)** Comparison in the magnetization as a function of temperature $\mu_oH \perp b$ and $\mu_oH//b$ at $\mu_oH$ = 0.1T and T=2-300K, **(d)** Curie Weiss analysis for $\mu_oH//b$ from $\mu_oH$ = 0.1 and T=2-300K, **(e)** Curie Weiss analysis $\mu_oH \perp b$ from $\mu_oH$ = 0.1 and T=2-300K, and **(f)** Ratio of magnetization $\mu_oH//b$ and $\mu_oH \perp b$ at $\mu_oH$ = 0.1T and T=2-300K displaying the anisotropy below $T_N$.

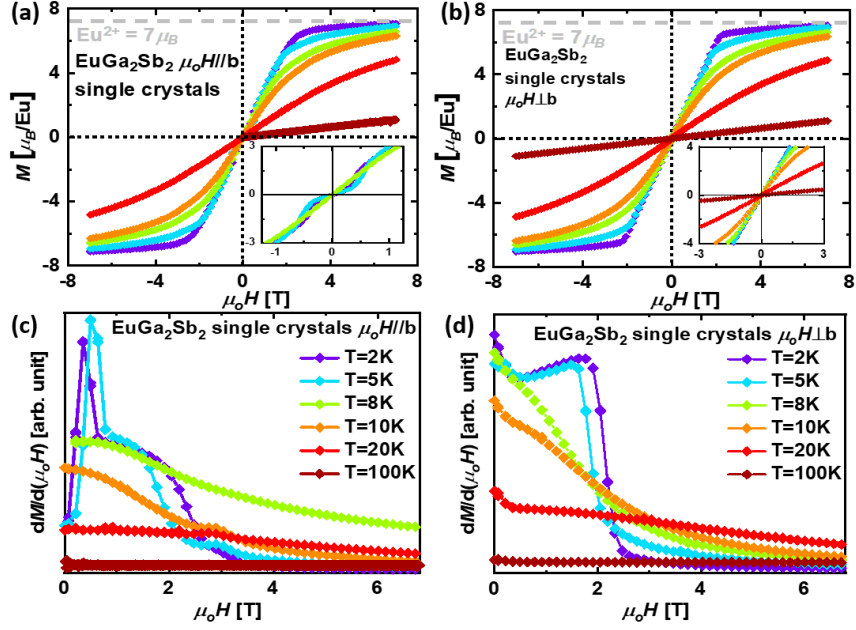

**Figure-5 (a)** Magnetization as a function of magnetic field with $\mu_oH//b$ from $\mu_oH$ = -7 to 7T and T=2, 5, 8, 10, 20, and 100 K. **(b)** Magnetization as a function of magnetic field with $\mu_oH\perp b$ from $\mu_oH$ = -7 to 7T and T=2, 5, 8, 10, 20, and 100 K, **(c)** derivative of magnetization over magnetic field as a function of magnetic field with $\mu_oH//b$ at T=2, 5, 8, 10, 20, and 100 K, and **(d)** derivative of magnetization over magnetic field as a function of magnetic field with $\mu_oH\perp b$ at T=2, 5, 8, 10, 20, and 100 K. There is a clear metamagnetic behavior below ~0.5 T.

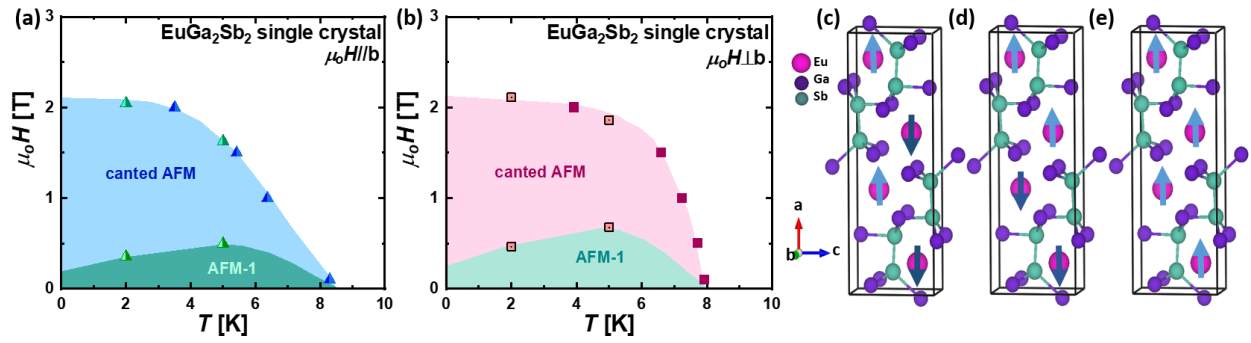

**Figure-6 (a)** Magnetic phase diagram of EuGa$_2$Sb$_2$ with $\mu_oH//b$. **(b)** Magnetic phase diagram of EuGa$_2$Sb$_2$ with $\mu_oH\perp b$. In both cases, there is an anisotropic AFM state at low fields, followed by a canted AFM state that continuously evolves to a field polarized state. **(c,d,e)** Possible magnetic structures for EuGa$_2$Sb$_2$. In (c,d), the chains are ferromagnetic along the chain direction (into the page), while in (e) the chains are antiferromagnetic along the chain direction.

| | Formula | EuGa$_2$Sb$_2$ |
|---|---|---|
| | Crystal system | Orthorhombic |
| | Space Group | *Pnma* (No. 62) |
| | a (Å) | 18.2201(4) |
| | b (Å) | 4.2987(1) |
| | c (Å) | 6.6994(1) |
| | V (Å$^3$) | 524.715(19) |
| | Z | 4 |
| | M/gmol$^{-1}$ | 534.93 |
| | $\rho_0$/gcm$^{-3}$ | 6.772 |
| | μ/mm$^{-1}$ | 31.895 |
| | Radiation | Mo K$\alpha$, λ= 0.71073 Å |
| | Temperature (K) | 90 K |
| | Reflections collected/number of parameters | 1481/86 |
| | Goodness-of-fit | 1.373 |
| | R[F]$^a$ | 0.0207 |
| | R$_w$(F$_o^2$)$^b$ | 0.00510 |

$^a$ R(F) = Σ||F$_o$| - |F$_c$||/Σ|F$_o$|
$^b$ R$_w$(F$_o^2$) = [Σw(F$_o^2$ - F$_c^2$)$^2$/Σw(F$_o^2$)$^2$]$^{1/2}$

| | Occ. | Wyckoff Positions | x (Å) | y (Å) | z (Å) | U$_{eq}$ (Å$^2$) |
|---|---|---|---|---|---|---|
| Eu | 1 | 4c | 0.3847(1) | 0.7500 | 0.7802(1) | 0.003(1) |
| Ga1 | 1 | 4c | 0.4293(1) | 0.2500 | 0.4418(1) | 0.004(1) |
| Ga2 | 1 | 4c | 0.2921(1) | 0.2500 | 0.4636(1) | 0.004(1) |
| Sb1 | 1 | 4c | 0.2872(1) | 0.2500 | 0.0640(1) | 0.003(1) |
| Sb2 | 1 | 4c | 0.4694(1) | -0.2500 | 0.2315(1) | 0.003(1) |

| | U(1,1) | U(2,2) | U(3,3) | U(1,2) | U(1,3) | U(2,3) |
|---|---|---|---|---|---|---|
| Eu | 0.00349 | 0.00349 | 0.00325 | 0 | 0.00002 | 0 |
| Ga1 | 0.00416 | 0.00369 | 0.00444 | 0 | 0.00001 | 0 |
| Ga2 | 0.00381 | 0.00367 | 0.00343 | 0 | 0.00001 | 0 |
| Sb1 | 0.00278 | 0.00282 | 0.00269 | 0 | 0.00015 | 0 |
| Sb2 | 0.00347 | 0.00321 | 0.00351 | 0 | -0.00026 | 0 |

**Table 1.** Crystallographic parameters of the SXRD for EuGa$_2$Sb$_2$.

| $s_{D1}$ (oscillator strength/formula unit) | $S_{D2}$ (oscillator strength/formula unit) | $\theta_{D1}$ (K) | $\Theta_{D2}$ (K) |
|---|---|---|---|
| 3.2(2) | 2.05(9) | 293(15) | 121(2) |

**Table 2.** Fitting parameters to the Cp/T as a function of T for EuGa$_2$Sb$_2$ to extract the phonon contribution.

| EuGa$_2$Sb$_2$ | $\mu_o H // b$ | $\mu_o H \perp b$ |
|---|---|---|
| Range [K] | 50-300 | 50-300 |
| C [emu.K.(mol Eu)$^{-1}$.Oe$^{-1}$] | 8.33 | 8.31 |
| θ [K] | 5.81 | 5.99 |
| $p_{eff}$ [$\mu_B$] | 8.16 | 8.15 |

**Table 3.** Fitting parameters obtained by Curie Weiss analysis from magnetization data of EuGa$_2$Sb$_2$.